\documentclass[manuscript,screen]{acmart}

\AtBeginDocument{%
  \providecommand\BibTeX{{%
    \normalfont B\kern-0.5em{\scshape i\kern-0.25em b}\kern-0.8em\TeX}}}

\setcopyright{acmlicensed}
\copyrightyear{2025}
\acmYear{2025}

\usepackage{todonotes}
\usepackage{soul}  % For highlighting text with \hl

\setcopyright{none}

\begin{document}

\title[vashTimer: Multimodal Presentation Timer Mobile App]{vashTimer: A Multi-Purpose, Multimodal Mobile App For Maintaining Passage of Time by means of Visual, Auditory, Speech, and Haptic Alerts}

\author{Aziz Zeidieh}
\orcid{0009-0000-9334-8660}
\affiliation{%
 \department{Informatics}
  \institution{University of Illinois Urbana-Champaign}
  \city{Champaign}
  \state{Illinois}
  \country{USA}
  \postcode{61820}
}
\email{azeidi2@illinois.edu}

\author{Sanchita S. Kamath}
\orcid{0000-0001-6469-0360}
\affiliation{%
 \department{School of Information Sciences}
  \institution{University of Illinois Urbana-Champaign}
  \city{Champaign}
  \state{Illinois}
  \country{USA}
  \postcode{61820}
  }
\email{ssk11@illinois.edu}

\author{JooYoung Seo}
\orcid{0000-0002-4064-6012}
\affiliation{%
 \department{School of Information Sciences}
  \institution{University of Illinois Urbana-Champaign}
  \city{Champaign}
  \state{Illinois}
  \country{USA}
  \postcode{61820}
  }
\email{jseo1005@illinois.edu}

\renewcommand{\shortauthors}{Zeidieh et al.}

\begin{abstract}
  Effective time management during presentations is challenging, particularly for Blind and Low-Vision (BLV) individuals, as existing tools often lack accessibility and multimodal feedback. 
To address this gap, we developed \textit{\textbf{vashTimer}}: a free, open-source, and accessible iOS application. 
This paper demonstrates the design and functionality of \textit{\textbf{vashTimer}}, which provides presenters with a robust tool for temporal awareness. The application delivers highly customizable alerts across four distinct modalities—visual, auditory, speech, and haptic—and supports multiple configurable intervals within a single session. By offering a flexible and non-intrusive time management solution, \textit{\textbf{vashTimer}} empowers presenters of all visual abilities.
The implications of this work extend beyond public speaking to any professional, such as a clinical therapist, who requires discreet temporal cues, fostering greater independence and focus for a wide range of users. 
This demonstration serves as the foundation for a planned formal user evaluation.
\end{abstract}





\maketitle

\section{Introduction}
\label{sec:introduction}

% 1: research problem.
Effective time management is a critical skill in many professional and academic settings. This principle is especially salient in the context of the oral presentation, a practice that requires communicating organized information to an audience under temporal constraints.
Conference presentations, specifically those attended in person, have been found to increase the serendipitous dissemination of presented works \cite{teplitskiyIntentionalSerendipitousDiffusion2024}. 
Therefore, improving one's presentation delivery is crucial, and effective time management is a key component of that success.
However, keeping track of time can be generally distracting. For blind and low-vision (BLV) presenters, this challenge is amplified, as they often must rely on others for time cues during their presentation.

% 2: prior studies that have addressed the problem.
Prior work on presentation time management has explored technological interventions ranging from mobile applications for section-based timing \cite{saketTalkZonesSectionbasedTime2014} and discrete haptic notification systems \cite{tamDesignFieldObservation2013} to empirical comparisons of different alert modalities \cite{gilASSESSINGINFLUENCEUSING2020}.
% 3: deficiencies in the studies.
While pivotal, these prior works have notable limitations. Existing solutions tend to focus on one or two feedback channels, neglecting a comprehensive approach that integrates visual, auditory, speech, and haptic modalities. A significant gap is the lack of consideration for accessibility, as these systems were not designed to address the specific needs of BLV presenters. Moreover, the resulting systems are typically confined to research contexts and are not made freely and publicly available as open-source tools for widespread community adoption.

% 4: Advance the significance of the study.
Therefore, the significance of our work extends to several audiences who require nuanced temporal awareness. The system we propose directly benefits both sighted and BLV presenters by offering customizable control over how they are alerted at given intervals during a timed presentation. Beyond public speaking, the utility of such a tool extends to clinical practitioners; for instance, massage therapists could receive discreet alerts to manage session intervals for switching focus areas or techniques. Ultimately, this contribution is relevant for any professional whose session has a defined duration, providing a flexible means to be aware of set intervals without disrupting the primary task at hand.

In this work, we set out to ask the following research questions.
\begin{itemize}
\item[\textbf{RQ1:}] How can multiple alert modalities including visual, auditory, speech, and haptics be leveraged to facilitate non-intrusive and informative notification to presenters of all visual abilities?
\item[\textbf{RQ2:}] How can multiple reminder intervals be implemented to provide more temporal awareness throughout the span of a timed presentation? 
\end{itemize}

Driven by these research questions, we developed \textit{\textbf{vashTimer}}; an open-source, accessible iOS application for flexible time management. The application offers deep customizability through four distinct alert modalities, visual, auditory, speech, and haptics \textbf{(VASH)}, which can be individually toggled to suit user preference or accessibility needs. To enhance its utility, \textit{\textbf{vashTimer}} allows users to define up to three alert intervals within a timed session and save their configurations as named presets for convenient reuse. 

% Looking ahead, we will review related work in Section~\ref{sec:related_work}, detail our design process and goals in Section~\ref{sec:design_procedures_and_goals}, present the system architecture in Section~\ref{sec:system_implementation}, and conclude with a discussion of future work in Section~\ref{sec:discussion}.
\section{Related Work}
\label{sec:related_work}

A significant body of research has focused on improving time management for oral presentations. Studies have explored custom hardware, such as the wrist-worn haptic system by \citet{tamDesignFieldObservation2013}, which provided a private communication channel between a speaker and a session chair to improve timing awareness. Other work has investigated the use of software, like the \textit{TalkZones} mobile application, which allows presenters to set time targets for sections of their talk, significantly reducing time overruns \cite{saketTalkZonesSectionbasedTime2014}. Further research directly compared different timekeeping devices, finding that while a vibrating remote did not have a statistically significant impact on timing adjustment, it did result in the best overall time adherence compared to a visual timer app \cite{gilASSESSINGINFLUENCEUSING2020}. Complementing these academic prototypes, various commercial applications such as CueTimer and Remote-Controlled Countdown Timer offer solutions, though they often lack open-source availability and deep customizability \cite{CueTimer2025, RemotecontrolledCountdownTimer}.
The design of our system is heavily informed by extensive research into the efficacy of haptic and multimodal feedback systems. Haptic communication, which focuses on integrating the sense of touch into digital experiences, is a vast and growing field of research \cite{emamiSurveyHapticsCommunication2025}. Studies consistently demonstrate that tactile alerts can be more effective than purely visual or auditory ones; for example, tactile warnings have been shown to yield the fastest driver reaction times in collision prevention scenarios \cite{scottComparisonTactileVisual2008, stanleyHapticAuditoryCues2006}. Furthermore, research indicates that multimodal warnings—those combining auditory and tactile feedback—can reduce the number of missed alerts compared to using a single modality, highlighting the performance benefits of a multi-sensory approach \cite{geitnerComparisonAuditoryTactile2019}. This superiority of haptic and multimodal feedback has also been confirmed in contexts of long-term supervision, where they are more effective than visual or auditory alerts alone \cite{souzaEvaluationVisualAuditory2018}.
In addition to haptic feedback, our work incorporates auditory alerts, with a specific focus on the utility of speech. The choice between different types of sounds is not trivial, as they are processed differently by the brain and are suited for different tasks. Research by \citet{glatzUseRightSound2018} used EEG/ERP methods to evaluate auditory notifications and found that while auditory icons are effective for communicating contextual information, verbal commands are more readily recognized by the brain as relevant targets for urgent requests. This distinction supports our inclusion of a speech-based modality, which can deliver clear, unambiguous information that a simple tone or beep cannot convey, making it ideal for distinct time-based alerts.

Finally, our design is grounded in principles of accessibility, particularly for BLV individuals. Non-visual modalities are not merely an alternative but are essential for creating effective assistive technologies. Research has shown that enriching a smartphone GPS map with haptic and audio hints can facilitate cognitive mapping and increase environmental awareness for BLV travelers, confirming the validity of this approach \cite{paratoreExploitingHapticAudio2024}. 
% FIND SOMETHING BETTER HERE THAN CLEW
The iOS platform, in particular, has been used to develop robust accessibility tools, such as the Clew Maps application for indoor navigation, demonstrating its suitability for creating solutions aimed at the BLV community \cite{abbotClewMapsCloudBased}. 
By integrating multiple non-visual feedback channels, our work aims to provide an inclusive and empowering tool for presenters of all visual abilities.
\section{Design Procedures and Goals}
\label{sec:design_procedures_and_goals}

\subsection{Design Procedures}
\label{subsec:design_procedures}

Over a three-week period starting in May 2025, \textit{\textbf{vashTimer}} was developed using the rapid iterative testing and evaluation (RITE) method \cite{medlock17ChapterRapid2005}. The process centered on an interdependent collaboration \cite{bennettInterdependenceFrameAssistive2018} between the BLV project lead (R1) and a non-BLV co-designer (R2), who worked together daily to iterate on and test \textit{\textbf{vashTimer}} on iPhone 14 Pro Max devices. As part of this daily cycle, R2 offered insights on the system's intuitive functionality and code modularity. In parallel, an BLV HCI researcher (R3), a screen reader user with 24 years of assistive technology experience, provided weekly end-user feedback from an iPhone 16e, guiding the project to the final version we present.

\subsection{Design Goals}
\label{sec:design_goals}

Development and design of \textit{\textbf{vashTimer}} was guided by the following design goals:

% For DGs, it looks much better than before. If you would like to improve it further, I suggest you talk about how we came up with each DGs. Currently, all of the DGs are supported by other related work. This is good, but  reviewers may want to hear more about "We" as co-designers reached those agreements. Say, for example, feedback from team members or our internal test reaveld revealed xxx. And this also aligns with previous work yy, zz. You may want to use citations once we provide our own design needs drived from our co-design sessions.

\begin{itemize}
\item[\textbf{DG1:}] \textit{\textbf{The system must provide an accessible, natively integrated interface to ensure equitable use by both screen reader and non-screen reader users.}} Grounded in ability-based design principles \cite{wobbrockAbilityBasedDesignConcept2011} and reinforced by our mixed-ability team, we aimed to create the system to be usable for presenters of all visual abilities.
\item[\textbf{DG2:}] \textit{\textbf{The system must support configurable mid-timer reminders to enhance the user's awareness of time progression during an activity.}} Cognitive–psychology studies indicate that periodic reminders reduce “task-time blurring” and help users maintain temporal awareness; using haptic or audio prompts at intervals could demonstrate improvements in user time perception accuracy \cite{NotificationsAwarenessField}.
\item[\textbf{DG3:}] \textit{\textbf{The system must offer flexible, user-selectable alert modalities (visual, auditory, speech, and haptic) to suit diverse contexts and preferences.}} 
Inspired by internal feedback and supported by research, we aimed to introduce multiple alert modalities. This allows presenters to choose their preferred, minimally distracting notification method while benefiting from the improved reaction times associated with multimodal (audio, haptic, and visual) stimuli \cite{yoshidaExploringHumanResponse2023}.
\item[\textbf{DG4:}] \textit{\textbf{The system must allow users to save and reuse custom timer configurations as presets to improve efficiency and streamline repeated tasks.}} Usability research emphasizes that providing presets for recurring configurations significantly reduces setup effort and cognitive load \cite{jolaoso2015taskambient, villalobos-zunigaAppsThatMotivate2020}. This observation was further solidified by internal input from our team who sought reusable timer presets for common timed tasks. 
\item[\textbf{DG5:}] \textit{\textbf{The system must be built upon a modular architecture to ensure long-term maintainability and facilitate future feature integration.}} The benefits of modular architecture in iOS development are well-documented: frameworks decomposing timing, alert, and UI components into independent modules lead to improved testability, faster builds, and easier extension, in line with proven engineering practices \cite{koushikCUSTOMCOMPONENTFRAMEWORK2024, vo2019application}.
\end{itemize}
\section{System Implementation}
\label{sec:system_implementation}

To address our design goals and research questions, we developed vashTimer, a multipurpose timer application that provides multiple alert modalities, visual, auditory, speech, and haptics (VASH).  

\subsection{Architecture}
\label{subsec:system_architecture}

The system architecture of \textbf{\textit{vashTimer}} is engineered around the CLEAN Architecture pattern and SOLID principles to create a robust, maintainable, and extensible application (\textit{DG5}). This design separates the system into four distinct layers: Presentation (UI), Domain (business logic), Data (services and persistence), and Core (dependency injection). This separation of concerns is fundamental to achieving the project's design goals.
The Domain Layer serves as the core of the application, encapsulating the essential business logic and rules, independent of any external frameworks. Key components within this layer directly address our design goals. For instance, the \textit{PresetUseCase} and its corresponding \textit{PresetRepositoryProtocol} provide the logic for creating, saving, and managing user-defined timer configurations, directly fulfilling the requirement for reusable presets (\textit{DG4}). Similarly, the \textit{ReminderIntervalsUseCase} contains the unified logic that allows for precise, configurable mid-timer reminders, satisfying the need for enhanced time awareness during an activity (\textit{DG2}).
Flexibility in alert modalities (\textit{DG3}) is achieved through a modular architecture implemented across multiple layers. In the Domain layer, the \textit{AppSettings} entity defines independent boolean properties for visual, auditory, speech, and haptic alerts. The Data Layer implements the concrete services that produce these alerts, such as the \textit{AudioService} and \textit{HapticService}. The \textit{TimerUseCase} in the Domain layer then triggers these services based on the user's saved preferences, ensuring that only the desired alert types are activated.
The Presentation Layer, built with SwiftUI, is responsible for the user interface and interaction. Its design ensures that all functionality is accessible. By using native SwiftUI components, such as individual \textit{Toggle} controls for each alert modality, the interface is inherently compatible with screen readers like VoiceOver, providing an equitable experience for all users (\textit{DG1}).
Finally, the strict adherence to a protocol-oriented design and dependency injection facilitates a highly modular system (\textit{DG5}). High-level modules depend on abstractions (protocols) defined in the Domain layer, not on concrete implementations in the Data layer.
This structure not only makes the system highly testable but also allowed for seamless extension, such as the development of a companion Apple Watch application, with its complex data synchronization logic managed entirely by a dedicated \textit{WatchConnectivityService} component on each device.

\subsection{User Interaction and Experience}
\begin{figure}[htbp]
  \centering
  \includegraphics[width=0.3\textwidth]{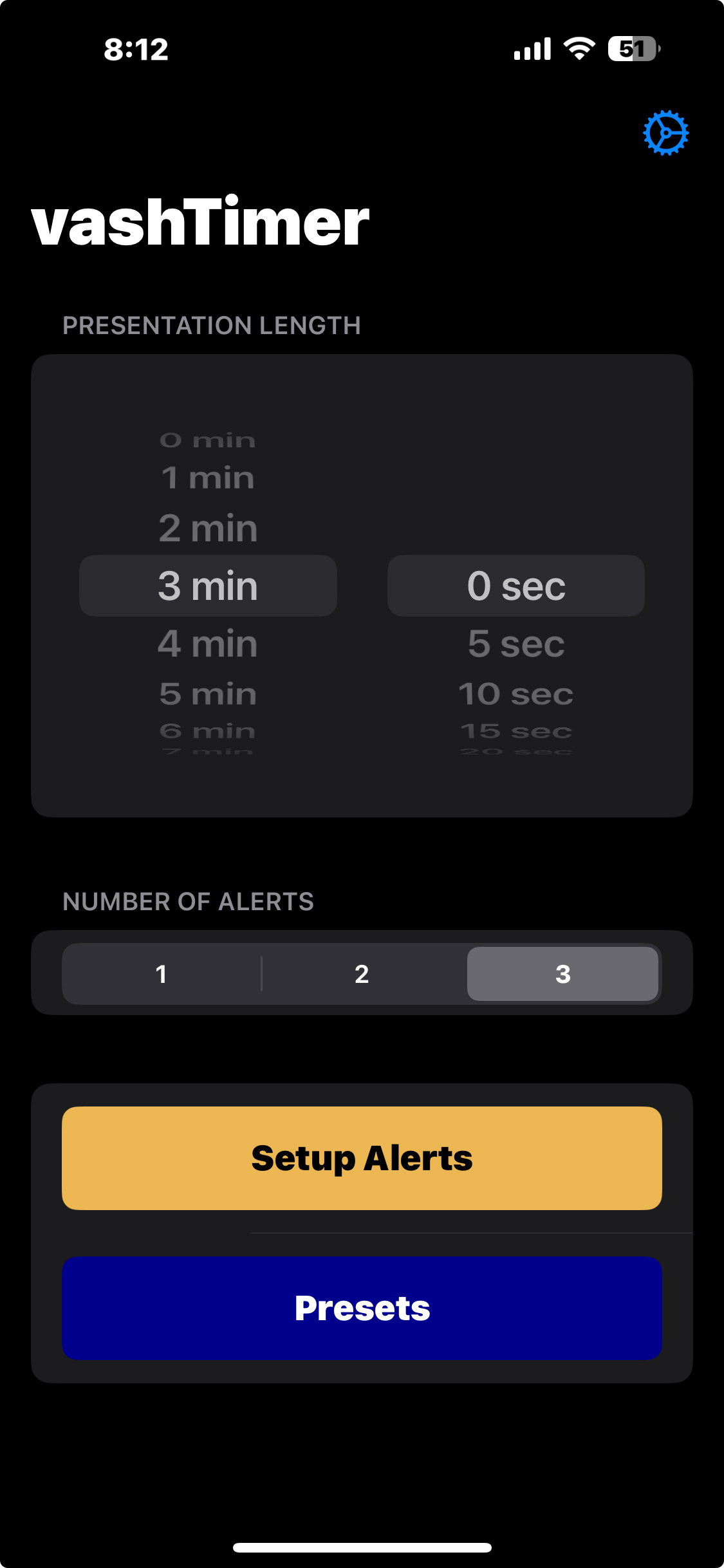}\Description{First screenshot: A smartphone screen shows an app interface named vashTimer. There is a section titled "PRESENTATION LENGTH." This section has a scrollable picker with minutes on the left and seconds on the right. The selected time is 3 minutes and 0 seconds. Below that, there is a section labeled "NUMBER OF ALERTS" with three options: 1, 2, and 3. The option "3" is currently selected. At the bottom, there are two large buttons. The top button is yellow with black text that says "Setup Alerts." The bottom button is blue with white text that says "Presets."}
  \includegraphics[width=0.3\textwidth]{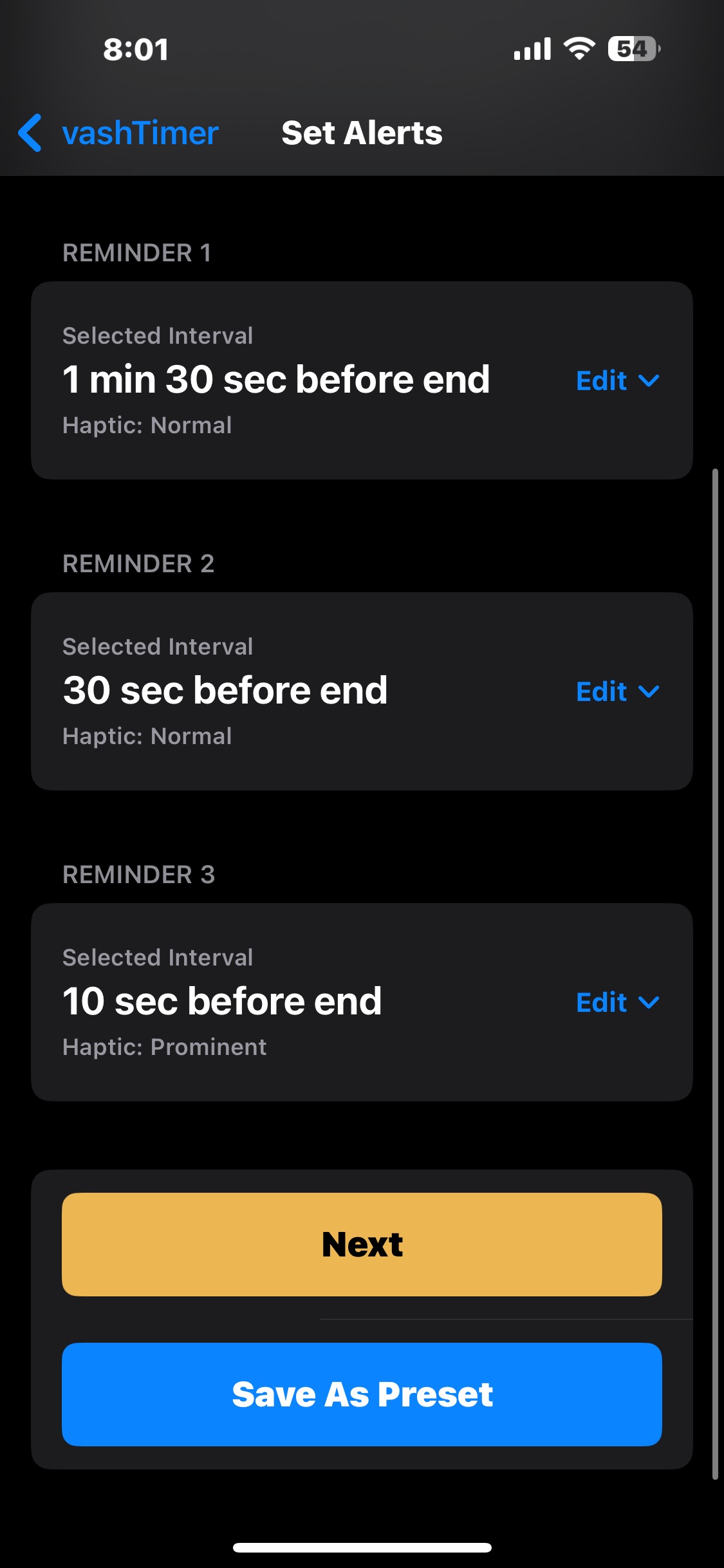}\Description{Second screenshot: the “Set Alerts” page is shown. There are three reminders set: Reminder 1 at 1 min 30 sec before end with Haptic set to Normal; Reminder 2 at 30 sec before end with Haptic set to Normal; and Reminder 3 at 10 sec before end with Haptic set to Prominent. For each reminder, there is an "Edit" option in blue on the right. Below the reminders, there are two buttons: A yellow button labeled "Next" and a blue button labeled "Save As Preset"}
  \includegraphics[width=0.3\textwidth]{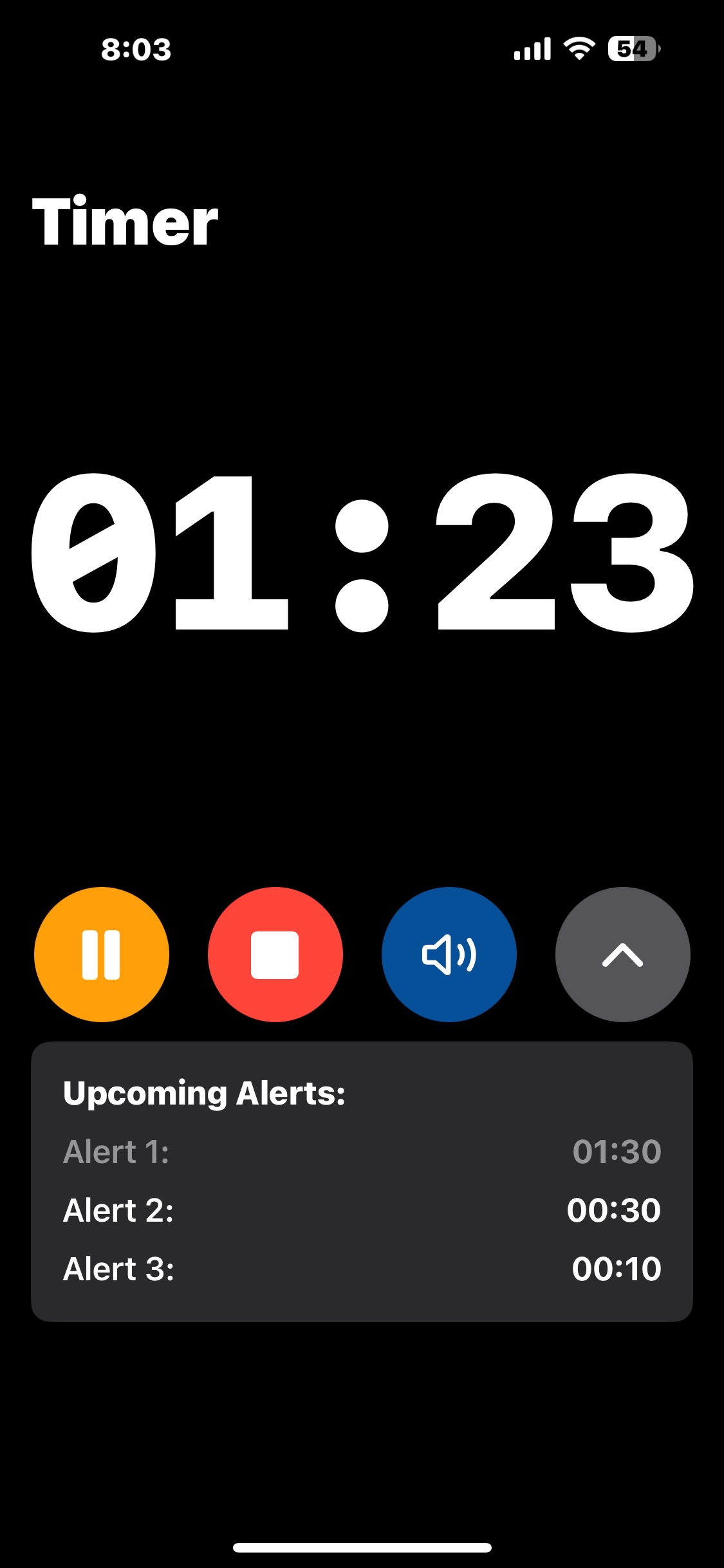}\Description{third screenshot: the timer screen is shown counting down at 1 minute and 23 seconds. Below the timer, there are four circular buttons: an orange pause button, a red stop button, a blue sound icon button, and a gray upward arrow button. Underneath these buttons, there is a section titled "Upcoming Alerts:" with three alerts listed, Alert 1 at 01:30 (in gray text), Alert 2 at 00:30 (in bold white text), and Alert 3 at 00:10 (in bold white text)}
  \caption{Duration picker, reminder interval setup, and timer view.}
  \label{fig:ui}
\end{figure}

The user interface is designed to be intuitive and accessible, providing a functional utility for both non-screen reader and screen reader users.  The user’s journey is centered around creating, saving, and using highly customized timer configurations.
The primary interaction begins with setting a total presentation duration as well as number of reminder intervals. From there, the user navigates to the \textit{Alert Setup View}, a unified interface for configuring all reminders.  Here, users can set up to three mid-presentation alerts with 5-second precision using minute and second pickers, allowing for fine-tuned awareness of time progression and addressing our second research question. The interface intelligently validates inputs to prevent errors and provides percentage-based defaults that adapt to the total duration. 
A key feature which addresses our first research question is the modular alert system, which allows users to select their preferred feedback modalities. Within the \textit{Settings View}, users are presented with four distinct toggles: \textbf{Visual}, \textbf{Auditory}, \textbf{Speech}, and \textbf{Haptic}.  Any combination of these alerts can be enabled, suiting diverse contexts and accessibility needs. 
Once a configuration is set, it can be saved as a named, reusable preset, streamlining the setup for recurring tasks.  When a timer is active, the \textit{Timer View} displays the time progression, which the user can toggle between a traditional countdown and a count-up "elapsed time" view from the \textit{Settings View}. Throughout this process, the Presentation Layer remains simple and declarative, delegating all logic to the Domain Layer to ensure a clean and accessible user experience.
\section{Future Direction}
\label{sec:discussion}

Informed by our own needs to support time keeping during public speaking through multiple alert modalities, our mixed-visual-ability team co-designed and co-developed \textit{\textbf{vashTimer}}, a free, open-source, and accessible tool that supports visual, auditory, spoken, and haptic alerts.
The design of \textit{\textbf{vashTimer}} holds broad implications for a wide range of people, including public speakers, clinical massage therapists, and any individual who requires granular temporal awareness for a timed activity. However, we acknowledge that \textit{\textbf{vashTimer}} is not without limitations.

A significant technical challenge stems from the constraints of Apple's WatchOS. To the best of our knowledge, the operating system does not permit a third-party application to generate custom haptic patterns while the device's screen is inactive, such as when a user's wrist is down. To circumvent this, we implemented a workaround that leverages a sequence of precisely timed local push notifications to simulate haptic alerts. However, this solution is not ideal, as it offers less temporal fidelity and control than native haptic playback and is dependent on the system's notification delivery timing.
While we cannot quantitatively measure the efficacy of our tool at this juncture, we believe its design holds broad implications.
A critical next step is to conduct a formal, participant-based study to rigorously evaluate vashTimer and the broad implications of its usability and effectiveness.

%Looking forward, a critical next step is to conduct a formal, participant-based study to rigorously evaluate \textit{\textbf{vashTimer}}. 
%Such a study would allow us to assess the application's usability and accessibility, and to compare its effectiveness against other offerings.

\begin{acks}
  The authors wish to express their gratitude for the support and contributions that have made this research possible. Detailed acknowledgments and disclosures of funding sources and collaborations will be provided upon completion of the double-blind review process to maintain the integrity of the review.
\end{acks}

\bibliographystyle{ACM-Reference-Format}
\bibliography{references/vash}

\end{document}